\documentclass{article}

\usepackage{graphicx}

\begin{document}

\def\be{\begin{equation}}
\def\ee#1{\label{#1}\end{equation}}

\title{Irreversible Processes in Inflationary Cosmological Models}
\author{G. M. Kremer\thanks{kremer@fisica.ufpr.br}  $\,$ and 
F. P. Devecchi\thanks{devecchi@fisica.ufpr.br}\\
Departamento de F\'\i sica, Universidade Federal do Paran\'a\\
Caixa Postal 19044, 81531-990, Curitiba, Brazil
}
\maketitle
\date{}
\begin{abstract}
By using the thermodynamic theory of irreversible processes
and  Einstein general relativity, a cosmological model 
is proposed where the early universe is 
considered as a mixture of a scalar field with a matter field.  
The scalar field 
refers to the inflaton while the matter field to the classical particles. 
The irreversibility is related to a particle  production process
at the expense of the gravitational energy and of the inflaton energy.
The particle production process is represented by a non-equilibrium
pressure in the energy-momentum tensor. The non-equilibrium pressure is
proportional to the Hubble parameter and its proportionality factor
is identified with the coefficient of bulk viscosity. The dynamic equations 
of the inflaton and the Einstein field equations determine the time 
evolution of the cosmic scale factor, 
the Hubble parameter, the acceleration and of the energy densities of the 
inflaton 
and matter. Among other results it is shown that in some regimes the 
acceleration 
is positive which simulates an inflation. Moreover, the acceleration 
decreases and 
tends to zero in the instant of time where the energy density of matter 
attains 
its maximum value.

\end{abstract}

\noindent
PACS: 98.80.-k; 98.80.Cq
\section {Introduction} 
 Cosmological models are among the most important formulations that can be
put under analysis
when we combine the thermodynamics of irreversible processes with Einstein's  
general
relativity~\cite{Wein,Mur,Bel,Gr,RPa,HC,HC1,Zim,Di,KD}. Using  these theories 
the different eras of our
 universe can be modeled; the subject  received great attention: the starting
point was to consider the hypothesis of homogeneity and isotropy in the form 
of the
 Robertson-Walker (RW) metric. More specifically, the  early universe physics 
is of great
 interest; the so-called inflationary formulations focus on that
era, when - according to these models - a rapid expansion of space-time
 would occur.
Using these ideas it is   possible to construct a non-equilibrium
scenario, where  a non-equilibrium pressure term is included in the 
energy-momentum
 tensor; this introduces the possibility of
creation of particles in the  early universe by means of the thermodynamic 
theory of irreversible processes.
In the 3+1 case the inclusion of this term  was proposed by
 Murphy \cite{Mur} who  
found   a solution that corresponds  to a simple expansion. 
In the inflation driven models a fundamental field, the inflaton, simulates
the early universe  with an
  equation of state that will adjust the form of the
potential and will affect  the temporal  evolution of
fundamental  quantities like the cosmic
scale factor and the universe's energy density.

In this work we start with the Einstein field equations, 
the inflaton dynamics (in a curved space-time), the conservation law 
for the energy-momentum tensor, the 
barotropic equations of state of the inflaton and matter fields. The
non-equilibrium pressure is considered as a
constitutive quantity - within the framework of ordinary 
(also known as first order Eckart) thermodynamic theory -  which is
proportional to the Hubble parameter and whose proportionality factor
is identified with the coefficient of bulk viscosity. From these equations 
we determine the time evolution of the cosmic scale factor, 
the Hubble parameter, the acceleration and of the energy densities of the 
inflaton 
and matter. The evolution equations for these quantities are function
of three parameters which are identified with  the cosmological constant and
two coefficients that appear in the barotropic equations of state for the 
pressures of the inflaton and of the matter.
It is shown that among several possible regimes the acceleration is found
to be a positive quantity which simulates an inflation. Moreover, 
in those regimes the acceleration decreases and 
tends to zero in the instant of time where the energy density of matter 
attains its maximum value.

The manuscript is structured as follows. In Section II we introduce 
the balance equations for the particle four-flow, energy-momentum tensor
and entropy four-flow and use the Gibbs equation  to 
identify the non-equilibrium pressure with the process of particle production. 
We discuss the inflaton equations in Section III and obtain the evolution 
equation
of the energy density of the inflaton as a function of the cosmic scale factor.
In Section IV we introduce the Einstein field equations and find  a 
second-order
differential equation for the evolution of the cosmic scale factor 
which is a function of three above mentioned parameters.
In Section V we search for  the evolution equations
of the cosmic scale factor, Hubble parameter, acceleration and 
of the density energies of the inflaton and matter for given values of the 
three  parameters and initial conditions for the
cosmic scale factor and Hubble parameter. Finally, in Section VI we consider
the non-equilibrium pressure as a variable within
the framework of extended 
(also known as causal or second-order) thermodynamic theory and discuss 
the behavior of the solutions.
Throughout this paper units have been chosen so that the speed of light
$c$ and Planck constant $\hbar$ are equal to one. 
\section{The Balance Equations}

According to the cosmological principle the universe is spatially 
homogeneous and isotropic so that it looks the same to all observers.
These assumptions imply that the universe can be described by the  
Robertson-Walker  metric whose line element 
in a space-time,  characterized
by the metric tensor $g_{\mu\nu}$ with signature $(+ \,-\,-\,-)$,  is given by
\be 
ds^2=dt^2-a(t)^2\left[{dr^2\over 1-kr^2}+r^2\left(d\theta^2+\sin^2\theta 
d\varphi^2\right)
\right].
\ee{1}
In the above equation $a(t)$ is the cosmic scale factor which is an unknown 
function of the time $t$ and $k$ may assume the values 1, 0, -1 that 
correspond to closed, flat and open universes, respectively.

We consider that the universe can be modeled as a mixture of two constituents,
namely a scalar field  and 
a matter field, in interaction via the gravitational field represented by the 
cosmic scale factor $a(t)$. There are other ways of modeling this 
interaction and for more details one is referred to Yokoyama and 
co-workers~\cite{Yo}, Berera~\cite{Be},
Billyard and Coley~\cite{BC} and the 
references therein. The scalar field refers to the so-called 
inflaton~\cite{AST,Linde}
and the matter field describes the particles classically. 
The mixture is characterized by the fields of particle four-flow $N^\mu$, 
energy-momentum tensor $T^{\mu\nu}$ and entropy four-flow $S^\mu$ whose 
balance equations read
\be
{N^\mu}_{;\mu}=n\Sigma,\qquad
{T^{\mu\nu}}_{;\nu}=0,\qquad
{S^\mu}_{;\mu}\geq0,\qquad
\ee{1a}
where $n$ denotes the particle number density while $\Sigma$ is the particle 
production rate.

In a homogeneous and isotropic universe the decompositions of the  
particle four-flow, energy-momentum tensor and entropy four-flow  
with respect to the four-velocity $U^\mu$ (with $U^\mu U_\mu=1$) read 
\be
N^\mu=nU^\mu,\qquad T^{\mu\nu}=(\rho+p+\varpi)U^\mu U^\nu-
(p+\varpi)g^{\mu\nu},\qquad S^\mu=nsU^\mu,
\ee{1b}
since in this case the stress tensor,
the heat flux and the entropy flux vanish. In (\ref{1b}) $\rho$, $p$ 
and $s$ represent the 
energy density, the  pressure and the entropy 
per particle of the mixture, respectively. The term $\varpi$ denotes the 
dynamic pressure which refers to a non-equilibrium pressure; this quantity
also represents  the production of particles phenomenologically~\cite{Z}. 
For this 
particular case the  Eckart and the Landau and Lifshitz  decompositions 
(see for example~\cite{CK1})  coincide.

Insertion of the representations (\ref{1b}) into the balance equations 
${N^\mu}_{;\mu}=
n\Sigma$, $U_\mu{T^{\mu\nu}}_{;\nu}=0$, and ${S^\mu}_{;\mu}\geq0$ lead to
\be
\dot n+n\Theta=n\Sigma,\qquad \dot \rho+(\rho+p+\varpi)\Theta=0,\qquad
n\dot s+ns\Sigma\geq0,
\ee{1c}
which represent the balance equations for the particle number density, 
energy density
and entropy density, respectively. In these equations we have introduced the 
usual notations: $\Theta\equiv {U^\mu}_{;\mu}$ for the expansion rate and 
$\dot n\equiv U^\mu n_{;\mu}$ for the covariant derivative along $U^\mu$.

In terms of the the covariant derivative along $U^\mu$ the Gibbs equation reads
\be
n\dot s={1\over T}\left(\dot\rho-{\rho+p\over n}\dot n\right),
\ee{1d}
where $T$ is the absolute temperature. If we eliminate from the Gibbs equation
the derivatives of the particle number density $\dot n$ and of the energy 
density 
$\dot \rho$ by using the two first balance equations given in (\ref{1d}) 
we get
\be
n\dot s+ns\Sigma={1\over T}\left(-\varpi\Theta-n\mu\Sigma\right)\geq0,
\ee{1e}
where $\mu=(\rho+p)/n-Ts$ is the chemical potential.

An adiabatic  process is characterized by the requirement that $\dot s=0$ so 
that 
we  can infer from (\ref{1e}) the inequality $ns\Sigma\geq0$. 
As was pointed out by Prigogine et 
al~\cite{Pri} the process of particle production leads to a non-symmetric 
relationship 
between space-time and matter since it refers to an irreversible process of 
particle production due to the gravitational energy. Moreover, we get from 
(\ref{1e}) when  $\dot s=0$ that
\be
\varpi=-(\rho+p){\Sigma\over\Theta}.
\ee{1f}
Hence, the dynamic pressure $\varpi$ can be  identified with the particle 
production
rate $\Sigma$. This result was first obtained, to the best of our knowledge, 
by Zimdahl~\cite{Z}.

We write the energy density  and the pressure of the mixture as a sum of  
two terms
that describe the scalar and the matter fields, e. g.,
\be
\rho=\rho_\phi+\rho_m,\qquad p=p_\phi+p_m,
\ee{1g}
where $\rho_\phi$, $ p_\phi$, and $\rho_m$, $p_m$ represent the 
energy density and the pressure of the inflaton and of the particles, 
respectively. In this case the energy-momentum tensor for the mixture  reads
\be
T^{\mu\nu}=(\rho_\phi+\rho_m+p_\phi+p_m+\varpi)U^\mu U^\nu-
(p_\phi+p_m+\varpi)g^{\mu\nu}.
\ee{2}

From this point on we shall consider a comoving frame where the four-velocity
is given by $(U^\mu)=(1,{\bf0})$. In this circumstance the expansion rate is 
given by $\Theta\equiv3H$ - where $H=\dot a/a$ is the Hubble parameter - and 
the
balance equation $U_\mu{T^{\mu\nu}}_{;\nu}=0$ can be written as
\be
\dot\rho_\phi+\dot\rho_m+3H(\rho_\phi+\rho_m+p_\phi+p_m+\varpi)=0.
\ee{3}
In a comoving frame the dot reduces to a differentiation with respect to the 
time.

\section{The Inflaton Equations}

Modern inflationary theories require the presence of the inflaton. This 
hypothetical particle is represented classically by a scalar field of the same 
name and the motivation behind these ideas can be found in the works
~\cite{AST,Linde,RP,Liddle,Gu}. 
The corresponding Lagrangian density (in a generic curved space-time)
for the scalar field $\phi(x^\mu)$ is written as
\be
{\cal L}={1\over 2}\partial_\mu\phi\partial^\mu\phi-V(\phi),
\ee{2a}
where $V(\phi)$ is  the potential density of the field. From the 
Euler-Lagrange equations
of motion one can obtain  the dynamics of a homogeneous scalar field 
\be
\ddot\phi+3H\dot\phi+{dV(\phi)\over d\phi}=0.
\ee{7}

If we consider the inflaton as a perfect fluid with an energy-momentum tensor
of the form $(T^{\mu\nu}_\phi)={\rm diag}(\rho_\phi, p_\phi, p_\phi, 
p_\phi)$, we can equate this formula to the expression coming from
Noether theorem
\be
T^{\mu\nu}_\phi=\partial^\mu\phi\partial^\nu\phi-{\cal L}g^{\mu\nu},
\ee{2b}
and obtain the  equations for the energy density 
and for the pressure  
in terms of the homogeneous scalar field, which are given by
\be
\rho_\phi={1\over 2}\dot\phi^2+V(\phi),\qquad
p_\phi={1\over 2}\dot\phi^2-V(\phi).
\ee{6}

We differentiate (\ref{6})$_1$ with respect to the time, make use of  
(\ref{7}) and get
the following  evolution equation for the energy density of the inflaton
\be
\dot\rho_\phi+3H(\rho_\phi+p_\phi)=0.
\ee{8}

Equations (\ref{3}) and (\ref{8}) show that the energy densities of the 
inflaton and of 
the matter decouple so that we can write
\be
\dot\rho_m+3H(\rho_m+p_m+\varpi)=0.
\ee{3a}

Now we follow the work of Ratra and Peebles~\cite{RP} 
and suppose that the pressure of the inflaton  is connected with its energy 
density  according to the barotropic equation of state
\be
p_\phi=(\nu-1)\rho_\phi,\qquad \hbox{where}\qquad 0\leq\nu\leq1.
\ee{9}
As we said before, in this model the homogeneous scalar field does not 
interact with other non-gravitational fields~\cite{RP}.

We insert (\ref{9}) into (\ref{8}) and get by integration a relationship 
between the 
energy density and the cosmic scale factor 
\be
{\rho_\phi\over\rho_\phi^0}=\left({a_0\over a}\right)^{3\nu},
\ee{12}
where $\rho_\phi^0$ and $a_0$ are the values of the energy density of 
the inflaton
and of the cosmic scale factor at $t=0$ (by adjusting clocks), respectively. 
The relationship
between the potential density of the scalar field and the cosmic scale factor 
can be obtained
from (\ref{6}), (\ref{9}) and (\ref{12}) yielding
\be
V=\rho_\phi^0\left({2-\nu\over 2}\right)\left({a_0\over a}\right)^{3\nu}.
\ee{12a}
The above result is in agreement with the work of Ratra and Peebles~\cite{RP}
as expected.

Once the time evolution of the cosmic scale factor is known
it is possible to obtain 
from (\ref{12}) and (\ref{12a})  the time evolution of the energy density and 
of the 
potential density of the scalar field. The determination of the cosmic scale 
factor 
from the Einstein field equations  will be the subject of the next section.

\section{Einstein Field Equations}

From Einstein field equations
\be
R_{\mu\nu}-{1\over 2}Rg_{\mu\nu}=-{8\pi\over m_{\rm P}^2}T_{\mu\nu},
\ee{4a}
where $m_{\rm P}=1/\sqrt G$ denotes the Planck mass, $R_{\mu\nu}$ the Ricci 
tensor and $R$ the 
 curvature scalar, one can get (working with the RW metric)
a system of equations that reads
\be
{\ddot a\over a}=-{4\pi\over 3m_{\rm P}^2}(\rho_\phi+\rho_m+3p_\phi+3p_m+
3\varpi),
\ee{4}
\be
H^2+{k\over a^2}={8\pi\over 3m_{\rm P}^2}(\rho_\phi+\rho_m).
\ee{5}
The system of  equations (\ref{3}), (\ref{4}) and (\ref{5})  is not 
linearly independent, since the
differentiation of (\ref{5}) with respect to the time, taking into account
the conservation law (\ref{3}) leads to (\ref{4}).

Let us first analyze the case that correspond to a false vacuum where $\nu=0$ 
so that 
$p_\phi=-\rho_\phi=-\rho_\phi^0$.
In this case we have only the scalar field, i. e., the matter field is absent 
($\rho_m=0$) and it follows from (\ref{5})
\be
\left({\dot a\over a}\right)^2+{k\over a^2}={8\pi\rho_\phi^0\over 3m_{\rm P}^2}
\equiv\chi,
\ee{13}
where $\chi$ can be identified with the cosmological constant $\Lambda
\equiv3\chi$.
Equation (\ref{13}) leads to the de Sitter exponential solution 
$a= a_0\exp\left(\sqrt{\chi}\,t\right)$ if one neglects the term $k/a^2$
during the rapid expansion.

For $\nu\neq0$ we can get from (\ref{5}) together with (\ref{3}) 
\be
\dot H={k\over a^2}-{4\pi\over m_{\rm P}^2}(\rho_\phi+\rho_m+
p_\phi+p_m+\varpi).
\ee{10a}

In order to determine the time evolution of the cosmic scale factor from 
(\ref{10a})
one has to know the relationship between the pressure $p_m$ and the energy 
density
$\rho_m$ of the matter and the constitutive equation for the dynamic 
pressure $\varpi$, since $p_\phi=(\nu-1)\rho_\phi$ and the energy density 
of the scalar field 
$\rho_\phi$ is given by (\ref{12}). Here we follow the works~\cite{Mur,KD} 
and write
\be
p_m=(\gamma-1)\rho_m,\qquad 1\leq\gamma\leq2,
\ee{10}
\be
\varpi=-3\eta H,\qquad \eta=\alpha(\rho_\phi+\rho_m).
\ee{11}
Equation (\ref{10}) is the  barotropic equation of state for the matter and 
some values
for $\gamma$ are: a) dust $\gamma=1$; b) radiation $\gamma=4/3$; c) 
non-relativistic matter
$\gamma=5/3$ and d) stiff matter (or Zel'dovich fluid~\cite{Gr}) $\gamma=2$. 
Equation 
(\ref{11}) relates the dynamic pressure $\varpi$ with the Hubble 
parameter $H$. The 
proportionality factor $\eta$ is the coefficient of bulk viscosity 
which is supposed to be
proportional to the energy density of the mixture $\rho=\rho_\phi
+\rho_m$.
Moreover, $\alpha$ is a constant.

The evolution equation for the cosmic scale factor is obtained from
(\ref{10a}) together with (\ref{12}),  (\ref{10}) and (\ref{11}) yielding
\be
\dot H={k\over a^2}-{3\over 2}\left[(\nu-\gamma)\chi
\left({a_0\over a}\right)^{3\nu}+(\gamma-3\alpha H) \left(H^2+{k\over 
a^2}\right)
\right].
\ee{14}
The above equation is a function of four parameters - namely $\nu$, $\chi$, 
$\gamma$ and $\alpha$ - for a given value of $k$. We can express one of these 
parameters as a function of the others. 
Indeed if
we consider that at $t=0$ (by adjusting clocks) we have $H(0)=H_0$, 
$\rho_m(0)=0$ and $a(0)=a_0$  it follows from (\ref{5}) and (\ref{14})
\be
H_0^2+{k\over a_0^2}=\chi,\qquad
\alpha={3\nu\chi a_0^2-2k\over 9H_0\chi a_0^2}.
\ee{15}

Now we make use of the relationships  (\ref{15})  and get from (\ref{14})
$$
\dot H={k\over a^2(\chi-k)}-{3\over 2}\left(\gamma -{3\nu\chi-2k\over 3\chi }
H\right)
\left(H^2+{k\over a^2(\chi-k)} \right)
$$
\be
-{3\over 2}(\nu-\gamma){\chi\over\chi-k}\left({1\over a}\right)^{3\nu},
\ee{16}
by introducing the dimensionless quantities: a) a new Hubble parameter 
$H\equiv H/H_0$; b) a new proper time $t\equiv t H_0$; c) a
new cosmological constant $\chi\equiv \chi a_0^2$ and d) a new cosmic
scale factor $a\equiv a/a_0$. Equation (\ref{16}) is a second-order 
differential
equation for the cosmic scale factor which is a function of three parameters:
the cosmological constant $\chi$, and the two coefficients that appear in the 
barotropic equations of state for the inflaton $\nu$ and matter $\gamma$.

In terms of the dimensionless quantities the energy densities read
\be
{\rho_\phi\over\rho_\phi^0}=\left({1\over a}\right)^{3\nu},\qquad
{\rho_m\over\rho_\phi^0}={\chi-k\over \chi}\left(H^2+{k\over 
a^2(\chi-k)} 
\right)-\left({1\over a}\right)^{3\nu}.
\ee{17}
The above equations connect the evolution of the energy densities with the 
evolution of the cosmic scale factor.
In the next section we analyze the time evolution of these quantities
for flat,
closed and open universes ($k=0, \pm 1$) and for given values of the 
parameters $\nu$, $\chi$ and  $\gamma$.

\section{Results and Discussions}
In order to obtain the time evolution of the cosmic scale factor and of the 
energy densities from  equations (\ref{16}) and (\ref{17}) one has to specify
two initial conditions, since (\ref{16}) is a second-order differential 
equation 
for $a(t)$. Here we choose that at the instant of time $t=0$
(by adjusting clocks) the energy density of the inflaton has its maximum
value while the energy density of the matter attains its minimum value, namely
$\rho_\phi(0)=\rho_\phi^0=1$ and $\rho_m(0)=0$. According to (\ref{17}) these
initial conditions are equivalent to fix the value $a(0)=1$ for the cosmic
scale factor and  $H(0)=1$ for the Hubble parameter.
There is still much freedom  to find the solutions, since (\ref{16}) and 
(\ref{17})
do depend on the three parameters $\nu$, $\chi$ and  $\gamma$ for each kind
of universe, i. e., closed, open and flat. Among several possible regimes 
which can be found by choosing different values for the parameters 
$\nu$, $\chi$ and  $\gamma$ we fix our attention to those values which simulate
an inflation - i. e., where the acceleration is a positive quantity - and where
the energy of the matter attains its maximum value at the instant of time
where the period of the inflation ends - which corresponds to a vanishing 
acceleration.
For a closed universe $(k=1)$ one set of values that satisfies the above 
mentioned conditions are $\nu=0.9$, $\gamma=1.9$ and $\chi=2.0$. In figures 
1 and
2 we have also chosen the same values for a flat $(k=0)$ and open $(k=-1)$ 
universes in order to compare the different solutions.

In Fig. 1 it is plotted the cosmic scale factor $a$, its velocity $\dot a$
and acceleration $\ddot a$ as a function of time for closed, flat and
open universes. We infer from the curves that  an open universe evolves more 
rapidly than the two other types of universe, while a closed universe evolves 
more
slowly. The same conclusion can be draw out for the acceleration, since the 
curves
show that the inflation period of an open universe is the largest following 
the 
flat and the closed universes. We have plotted in Fig. 2 the energy densities 
of the
inflaton and of the matter as function of time. We note from this figure that 
the 
energy density of the matter grows more rapidly with respect to the time for 
an open universe following the cases of flat and closed universes. Moreover, 
we can 
infer that the energy density of the inflaton decays more slowly with 
respect to the
time for a closed universe following the cases of flat and open universes. 
Once the maximum
of the energy density of the matter is attained other models should be 
considered for the universe, since the universe reheats and 
follows the standard big-bang model.

One may wonder from Fig. 2 why the sum of the energy densities 
of the inflaton and of the matter is not a constant. This can be explained
as follows: in the presence of gravitational 
fields the energy-momentum tensor of the matter alone 
does not lead to a conservation law (see for example
Landau and Lifshitz~\cite{LL} and Dirac~\cite{Dir}), 
since one has to include the energy-momentum pseudo tensor of 
the gravitational 
field $T^{\mu\nu}_G$ which is given by
\be
T^{\mu\nu}_G=
{m_{\rm P}^2\over 16\pi}\left[{g^{\mu\tau}\over\sqrt{g}}(
\Gamma^\nu_{\alpha\beta}-\delta_\beta^\nu\Gamma^\sigma_{\alpha\sigma})
{\partial g^{\alpha\beta}\sqrt{g}\over \partial x^\tau}-g^{\alpha\beta}
(\Gamma^\sigma_{\alpha\beta}\Gamma^\tau_{\sigma\tau}
-\Gamma^\rho_{\alpha\sigma}\Gamma^\sigma_{\beta\rho})g^{\mu\nu}\right],
\ee{18}
where $\Gamma^\rho_{\alpha\sigma}$ are Christoffel symbols and $g\equiv
-\det(g_{\mu\nu})$.

For a RW metric with $k=0$ we get 
\be
T^{00}_G=-{3m_{\rm P}^2\over 8\pi}\left({\dot a\over a}\right)^2,
\ee{19}
hence giving the "expected" relation 
$T^{00}_G+T^{00}_\phi+T^{00}_m=0$ thanks to $T^{00}_\phi=\rho_\phi$, 
$T^{00}_m=\rho_m$ and (\ref{5}). One may say that the energy density of the
matter increases at the expense of the inflaton and 
gravitational energies.
This result must be handled with care 
since the energy density of the gravitational field is an intrinsically 
non-covariant quantity~\cite{LL,Dir}.

\section{Final Remarks}

In this work we have  analyzed the 
irreversible processes within the framework of ordinary thermodynamics 
 instead of extended thermodynamics 
 (see for 
example~\cite{Bel,RPa,HC,HC1,Zim,Di}). The reason is that - apart from 
causality -
there exist some thermodynamic systems in which 
ordinary and extended thermodynamics lead to the same results,
 like  the problems concerning the 
propagation of low frequency waves in fluids and those related to the
flow and the heat transfer
in fluids in the continuum regime (see for example~\cite{MR}). 
When the behavior of
the thermodynamic fields are not smooth enough, there exist differences 
between the solutions of  ordinary and 
extended thermodynamics. Here we confirm that both descriptions can lead 
to the  same behavior of the solutions and for this purpose
let us proceed to analyze the same problem 
by using  extended thermodynamics.

In extended thermodynamics the dynamic pressure $\varpi$ is no more given by 
the constitutive equation (\ref{11}) but it is  a variable whose 
evolution equation in a linearized theory reads (see for example~\cite{CK1,MR})
\be
\varpi+\tau\dot\varpi=-3\eta H,
\ee{20}
where $\tau$ is a characteristic time.  The evolution equations 
for the heat flux and pressure tensor are not consider here, since 
we are dealing with a  homogeneous and isotropic universe.
Hence we have to solve the system of 
evolution equations for the cosmic scale factor (\ref{10a}) and for the dynamic
pressure (\ref{20}).

According to the kinetic theory of relativistic gases (see for 
example~\cite{CK1}) the
characteristic time which appears in the evolution equation for the 
dynamic pressure
is given by $\tau=\eta/p$, where $p$ is the pressure of the mixture. Hence the 
system of equations (\ref{10a}) and (\ref{20}) can be written in terms of 
dimensionless
quantities as
\be
\dot H-{k\over a^2(\chi-k)}+{3\over 2}\left[{(\nu-\gamma)\chi\over \chi-k }
\left({1\over a}\right)^{3\nu}+\gamma
\left(H^2+{k\over a^2(\chi-k)} \right)+\varpi\right]=0,
\ee{21}
$$
\alpha(\chi-k)\dot\varpi+\left[\varpi
+3\alpha H\left(H^2+{k\over a^2(\chi-k)} \right)\right]
\left[(\nu-\gamma)\chi\left({1\over a}\right)^{3\nu}\right.
$$
\be
\left.+(\gamma-1)(\chi-k)
\left(H^2+{k\over a^2(\chi-k)} \right)\right]=0,
\ee{22}
by introducing the dimensionless quantities $\alpha\equiv\alpha H_0$ and 
$\varpi\equiv
8\pi\varpi/(3m_{\rm P}^2H_0^2)$, apart from those defined in Sec. IV.

As we have pointed out in previous sections there exist several possible 
regimes that can be
found by solving the system of equations (\ref{21}) and (\ref{22}), since 
it is a function
of four parameters $\nu$, $\chi$, $\gamma$ and $\alpha$. If one chooses the 
following initial
conditions $a(0)=1$, $H(0)=1$ and $\dot\varpi(0)=0$ one may obtain for (say) 
$\nu= 0.9$, 
$\chi=2.0$, $\gamma=1.9$,  $\alpha=0.1$ and 
$k=1$ a solution for the fields in extended (causal)
thermodynamics 
which have the same behavior as those represented in Figs. 1 and 2 
by using ordinary (Eckart) thermodynamics.

\newpage

\begin{figure}
\begin{center}
\includegraphics[width=8cm]
{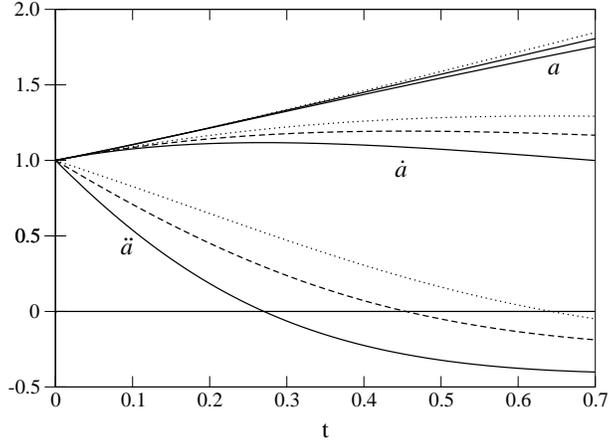}
\end{center}
\caption{Cosmic scale factor $a$, velocity $\dot a$ and acceleration $\ddot a$
vs time $t$ for open $k=-1$ (dot-line), flat $k=0$ (dash-line) 
and closed $k=1$ (straight-line) universes.}
\end{figure}

\vskip1cm
\begin{figure}
\begin{center}
\includegraphics[width=8cm]
{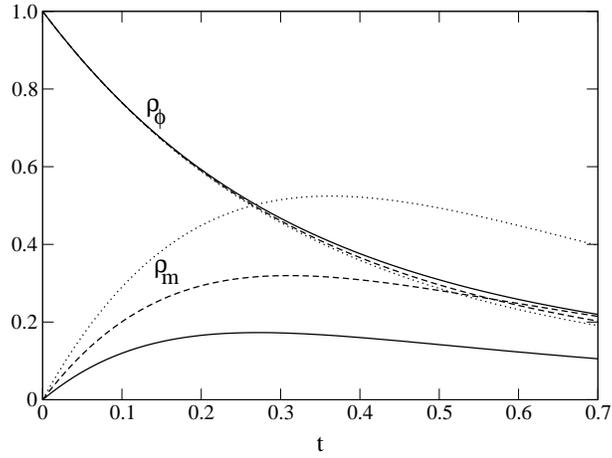} 
\end{center}
\caption{Inflaton energy density $\rho_\phi$ and matter energy density 
$\rho_m$ vs time $t$ for open $k=-1$ (dot-line), flat $k=0$ (dash-line) 
and closed $k=1$ (straight-line) universes.}
\end{figure}

\end{document}